# Superconducting phase diagram of multi-layer square-planar nickelates

Grace A. Pan[1]†, Dan Ferenc Segedin[1]†, Sophia F. R. TenHuisen[1,2], Lopa Bhatt[3], Harrison LaBollita[4], Abigail Y. Jiang[1,2], Qi Song[1], Ari B. Turkiewicz[1], Denitsa R. Baykusheva[1], Abhishek Nag[5], Stefano Agrestini[5], Ke-Jin Zhou[5], Jonathan Pelliciari[6], Valentina Bisogni[6], Hua Zhou[7], Mark P. M. Dean[8], Hanjong Paik[9], David A. Muller[3,10], Lena F. Kourkoutis[3,10], Charles M. Brooks[1], Matteo Mitrano[1], Antia S. Botana[4], Berit H. Goodge[11], and Julia A. Mundy[1,2]*

**Affiliations:**

[1]Department of Physics, Harvard University, Cambridge, 02138, USA.
[2]John A. Paulson School of Engineering and Applied Sciences, Harvard University, Cambridge, 02138, USA.
[3]School of Applied and Engineering Physics, Cornell University, Ithaca, 14853, USA.
[4]Department of Physics, Arizona State University, Tempe, 85287, USA.
[5]Diamond Light Source, Harwell Campus, Didcot OX11 0DE, UK
[6]National Synchrotron Light Source II, Brookhaven National Laboratory, Upton, 11973, USA.
[7]X-ray Science Division, Advanced Photon Source, Argonne National Laboratory, Lemont, IL 60439, USA
[8]Condensed Matter Physics and Materials Science Division, Brookhaven National Laboratory, Upton, New York 11973, USA.
[9]Platform for the Accelerated Realization, Analysis and Discovery of Interface Materials (PARADIM), Cornell University, Ithaca, 14853, USA.
[10]Kavli Institute at Cornell for Nanoscale Science, Cornell University, Ithaca, 14853, USA.
[11]Max Planck Institute for Chemical Physics of Solids, 01187, Dresden, Germany.

†These authors contributed equally to this work.

*Corresponding author. Email: mundy@fas.harvard.edu

**Abstract**: The discovery of superconductivity in square-planar nickelates has offered a rich materials platform to explore the origins of cuprate-like superconductivity. Experimental investigations however have largely been limited to the infinite-layer $R$NiO$_2$ ($R$=rare-earth) nickelates. Here, we construct a phase diagram of multi-layer square-planar Nd$_{n+1}$Ni$_n$O$_{2n+2}$ compounds and discover signatures of superconductivity for $n = 4 - 8$. Upon decreasing the dimensionality $n$, the superconducting anisotropy evolves due to 4$f$ electron effects, and electronic structure characteristics approach cuprate-like behavior. Magnetic fluctuations persist from within the superconducting regime and into the over-doped, non-superconducting regime. Remarkably, the superconducting regime overlaps with that of chemically-doped infinite-layer nickelates, demonstrating underlying commonalities and distinct differences across varying structural realizations of square-planar nickelates. Our work establishes this layered template for creating new nickel-based superconductors.

**Main Text:**

Following the discovery of high-temperature superconductivity in copper oxides (cuprates), there has been a longstanding search for similar superconductors in other materials systems (*1–4*). The nickel oxides (nickelates) have been an enduring candidate, as $Ni^{1+}$ and $Cu^{2+}$ have the same $3d^9$ electron count (*5, 6*). After decades of theoretical predictions and experimental attempts, superconductivity was eventually realized in thin films of Sr-doped $NdNiO_2$, commonly referred to as the "infinite-layer" nickelate (Fig. 1A) (*7*). Since then, the superconducting nickelates have expanded to include infinite-layer $RNiO_2$ with various rare-earth (La, Pr, Nd) and dopant (Sr, Ca, Eu) cations (*8–11*), the five-layer square-planar $Nd_6Ni_5O_{12}$ (*12*), and electronically distinct perovskite-like nickelates under applied pressure (*13*). Despite the close analogy between square-planar nickelates and cuprates, there are key distinctions between these compounds in their superconducting behavior, electronic structure, and correlated phases (*6, 14–22*). This positions the square-planar nickelates as a rich materials class in which to explore the mechanisms behind unconventional superconductivity.

Underlying a broad structural and chemical diversity, the hole-doped phase diagrams of the cuprates exhibit strong overlap in superconducting regions as well as in other proximate correlated features such as the strange metal, pseudogap, or Fermi-liquid like phases (*2, 23, 24*). This universal phenomenology in the cuprates has motivated the pursuit of unifying theoretical descriptions, including especially effective spin-1/2 models on an antiferromagnetic square lattice. At the same time, the parameterization of cuprates by their subtle differences such as apical oxygen-copper distances or the strength of orbital polarization may help to distill the essential materials ingredients for high-$T_c$ superconductivity (*2, 25*). Following this approach, a key question is which phenomenological features in the superconducting square-planar nickelates are similarly universal, and thus reflect the basic physics of a $d^9$ nickel-oxygen system, and which features are sensitive to the precise structural environment and can be judiciously tuned.

In this manuscript, we discover a superconducting regime in the multi-layer square-planar $Nd_{n+1}Ni_nO_{2n+2}$ compounds for $n$ = 4, 5, 6, 7, 8 (Fig. 2A), generalizing our ability to tailor the electronic properties in these chemically-undoped compounds through structural layering. Remarkably, this superconducting regime overlaps with that of the chemically-doped infinite-layer nickelates (*16, 26–28*). As we vary $n$, we observe that the electronic structure of lower $n$ compounds becomes more cuprate-like (*15, 29, 30*). Structural tuning, complementary to chemical doping, furthermore, enables unique access to the highly overdoped regime, where we observe the persistence of magnetic fluctuations in a metallic, non-superconducting phase. We also reveal features unique to the multi-layer $Nd_{n+1}Ni_nO_{2n+2}$ nickelates, namely a local lattice expansion of the atomic structure which may impact other properties including such magnetic fluctuations. Using atomically-precise design and synthesis, we establish the underlying universalities that manifest broadly in square-planar nickelates, with additional structurally-influenced effects which can template the construction of new superconducting and correlated materials.

**Structural description of the multi-layer square-planar nickelates:**

Multi-layer $Nd_{n+1}Ni_nO_{2n+2}$ compounds are structurally analogous to electron-doped cuprates (e.g. $Nd_{2-x}CeCuO_4$) which crystallize in the T' structural form. Written schematically as $(NdNiO_2)_n(NdO_2)$, $n$ layers of square-planar $NdNiO_2$ are hole doped by negatively charged $(NdO_2)^-$ fluorite layers (Fig. 1A) to an average nominal doping value of $1/n$ holes per $Ni^{1+}$ site (*6, 31–33*). These multi-layer square-planar nickelates can thus be directly mapped onto the phase diagram of the chemically-doped $R$NiO$_2$ nickelates in terms of nominal electron count, providing a structural route to traverse the correlated phase space (Fig. 1A).

We synthesized thin film Nd$_{n+1}$Ni$_n$O$_{2n+2}$ compounds for $n = 3 - 8$ following an improved version of the procedure we described previously (*12, 34–36*). Broadly, our procedure starts with the epitaxial synthesis of multi-layer perovskite-like Ruddlesden-Popper type Nd$_{n+1}$Ni$_n$O$_{3n+1}$ compounds. This is followed by a topochemical reduction process assisted by CaH$_2$ that removes the apical oxygen and reduces the valence from $d^{7+1/n}$ to $d^{9-1/n}$. (The Nd$_2$NiO$_4$ ($n = 1$) compound is known to be stable in just an octahedral, not square-planar coordination.) The large structural transformation between the Ruddlesden-Popper and multi-layer square-planar phases, combined with additional rare-earth spacer layers (Fig. 1A) in multi-layer nickelates, make this a particularly challenging synthesis (*34, 35, 37*). Nonetheless, sustained optimization of the existing growth parameters combined with new improvements to the synthesis process have enabled us to create the entire series for $n = 3 - 8$ (*36*) (Figs. S1-S5).

High-angle annular dark field scanning transmission electron microscopy (HAADF-STEM) of a representative $n = 6$ square-planar compound reveals a coherent atomic layered microstructure (Fig. 1B, C). The overall defect density is greatly reduced compared to initial reports of the Nd$_6$Ni$_5$O$_{12}$ phase (*12, 36*) (Fig. S5). Smaller field-of-view HAADF- and annular bright field (ABF) STEM images reveal an atomically sharp $n = 6$ block and nickel-oxygen coordination representative of the square-planar phase (*36*) (Fig. S6). Precise mapping of the intra-unit-cell lattice distances shows that the out-of-plane Nd-Nd distances vary through the $n$-layer block (Fig. 1B, Figs. S7-S9. (*36*)). The Nd-Nd distances are smaller in the interior of the $n$-layer block, approaching the $c$-axis lattice constant of undoped NdNiO$_2$. Toward the edge of the $n$-layer block close to the fluorite (NdO$_2$)$^-$ layer, the Nd-Nd distance increases. The local lattice expansion near the fluorite layers, also captured by structural relaxations in density functional theory (DFT) calculations (*36*), could suggest that other properties such as interlayer hopping or magnetic interactions may also be modified; we return to this point later in our discussion of magnetic excitations.

**Characteristics of the superconducting state:**

The principal result is the discovery of signatures of superconducting transitions in the $n = 4 - 7$ compounds with $T_{c, \text{ onset}}$ ranging from 9.9 K to 12.7 K (Fig. 2), and weak superconducting correlations in the $n = 8$ compound. This forms a region of superconducting behavior as a function of nominal nickel 3$d$ filling, or dimensionality $n$, corroborated by local spectroscopic measurements (Fig. S10). The maximal observed $T_{c, \text{ onset}}$ occurs at 12.9 K in $n = 6$, or a filling of $d^{8.83}$, approximately coincident in $d$ filling with maximal $T_c$ in recently optimized Nd$_{1-x}$Sr$_x$NiO$_2$ ($d^{8.825-8.85}$) (*16*). The $n = 4, 8$ compounds, corresponding to $d^{8.75}$ and $d^{8.875}$, respectively, show the weakest superconducting features, with superconducting downturns which lie on top of resistive upturn phases arising from localization or disorder-induced effects (Figs. 2B, S11, S12, (*36*)). These effects, along with the broad superconducting transitions, have been observed in a variety of superconducting nickelates (*7, 12, 38*) including in multi-layer nickelates La$_3$Ni$_2$O$_7$ (*39*) and Nd$_6$Ni$_5$O$_{12}$ (*38*). Hence the $n = 4, 8$ compounds likely represent the edges of the superconducting region accessible by dimensional doping.

The multi-layer square-planar nickelates offer a unique structural knob to tune the electronic properties: decreasing the layering order $n$ should decrease the electronic dimensionality and enhance the two-dimensional (2D) behavior of the superconducting confinement (*12, 40*). To this end, we interrogate the behavior of the superconducting transition in response to a magnetic field applied in- and out-of-plane of the film: a field applied perpendicular (parallel) to the *a-b* plane should maximally (minimally) suppress the superconducting transition in a 2D system due to orbital depairing effects (Fig. 3D) (*41*). Surprisingly, we see that in the most structurally 2D superconductor, $n = 4$, the superconducting anisotropy is minimal whereas the less structurally 2D superconductor, $n = 6$, exhibits a stronger superconducting anisotropy (Fig. 3A).

We attribute this reversal of the expected superconducting anisotropy to the effect of the 4*f* moments arising from the neodymium site (Fig. 3E), as originally proposed by Ref. (*29*). As suggested by neodymium- or cerium-containing electron-doped cuprates (*42*, *43*), the 4*f* moment on the rare-earth site forms an easy axis of enhanced magnetic permeability, enhancing the paramagnetic depairing along the in-plane direction, which is typically the direction of minimal orbital depairing. This effect appears enhanced with increased hole doping. Through angle-dependent magnetoresistive measurements (Fig. 3B) we observe a general reversal of the expected superconducting anisotropy as we move from $n = 8$ ($d^{8.875}$) to $n = 4$ ($d^{8.75}$) (Figs. 3B, 3F, S13), consistent with observations in increasingly Sr-doped $NdNiO_2$.

In $Nd_{n+1}Ni_nO_{2n+2}$ nickelates, increased hole doping is achieved by a net increase of neodymium content from the additional $(NdO_2)^-$ layers, suggesting that the permeability enhancement could trivially arise from having more neodymium or from a closer structural proximity of the 4*f* moments in the $(NdO_2)^-$ layers to the Ni-O planes. However, in Sr-doped $NdNiO_2$, increased hole doping requires the removal of neodymium. Hence, the strength of the enhanced magnetic permeability from the 4*f* moment contribution likely does not principally derive from the neodymium content despite arising from the neodymium site. This effect is also unrelated to the fragility of the superconducting phase: the "under-doped" $n = 8$ ($d^{8.875}$), which shows only weak superconducting correlations, shows weaker contributions from the 4*f* moments than the optimally and over-doped $n = 4 - 6$, which have sharper superconducting transitions. Finally, our ability to synthesize the heavily overdoped, non-superconducting $Nd_4Ni_3O_8$ ($n = 3$, $d^{8.67}$) compound allows us to rule out this effect as arising from non-superconducting impurity phases competing with the superconducting state. The non-superconducting $n = 3$ compound exhibits a very weak two-fold resistivity anisotropy which likely originates from localization effects, and importantly, is 90° out-of-phase with the feature generated by this paramagnetic depairing mechanism ((Fig. S14), (*36*)).

Combined, these observations suggest an underlying commonality within the nickelate phase diagram: the permeability enhancement by the neodymium 4*f* moment increases as we move toward increased hole doping in the superconducting region. In fact, the impact of the 4*f* moments is strong enough to completely obscure the effects of dimensional reduction by structural tuning. The $R_{n+1}Ni_nO_{2n+2}$ compounds are therefore optimal systems to study the interplay of superconducting dimensionality effects with rare-earth magnetism. By selectively tuning the rare-earth composition ($R$ = La, Pr, Nd) in $R_{n+1}Ni_nO_{2n+2}$ one could potentially isolate the impact of the 4*f* moments on the superconducting state and tune the magnetic interactions in lower dimensions (*29*, *44*). To this end, we have synthesized the $La_{n+1}Ni_nO_{2n+2}$ compounds, which have no *f* electrons but they are not superconducting to date, possibly due to strain-driven synthesis effects (*36*) (Figs. S17-S21).

**Electronic structure:**

A notable difference between cuprates and square-planar nickelates lies in their electronic structures (Fig. 4A). Cuprates can largely be described as charge-transfer insulators with an effective single band derived from *p-d* hybridized oxygen-copper states. In contrast, nickelates show reduced *p-d* oxygen-nickel hybridization and additional rare-earth 5*d* bands close to the Fermi level which may potentially self-dope the system, hybridize electronically with the nickel 3*d* bands, or even generate Kondo states (*14, 18, 45–47*). As dimensionality $n$ in the Nd$_{n+1}$Ni$_n$O$_{2n+2}$ nickelates is decreased however, DFT calculations find that lower $n$ nickelates should recover more cuprate-like features, with decreased 5*d*-3*d* (rare earth-nickel) hybridization and increased 2*p*-3*d* (oxygen-nickel) hybridization (Figs. 4B, S22, (*36*)). To experimentally assess the evolution of electronic structure characteristics with $n$ in the Nd$_{n+1}$Ni$_n$O$_{2n+2}$ compounds, we use resonant inelastic x-ray scattering (RIXS) at the nickel $L_3$ edge (Figs. S25, S26).

Representative RIXS spectra taken for the $n = 3$ and $n = 5$ compounds are shown in Fig. 4C. Both samples show orbital excitations consistent with square-planar oxygen coordination of nickel sites: a peak at 1.4 eV energy loss from $d_{xy}$ orbitals, a peak at 2 eV from $d_{xz}$ / $d_{yz}$ orbitals, and a higher energy shoulder from $d_{3z^2-r^2}$ orbitals as has been established in *R*NiO$_2$ (*30*). Above 5 eV, we observe a broad, higher energy feature which we attribute to Ni-O hybridization (*21, 48, 49*). This feature is stronger in the $n = 3$ than in the $n = 5$ compound, corroborating an Ni-O hybridization increase with decreasing $n$. At ~0.7 eV, we observe a lower energy shoulder that we attribute to Nd-Ni hybridization, which has also been observed in Refs. (*18, 30, 50*). This feature is weaker in the $n = 3$ compound than the $n = 5$ compound relative to the *dd* orbital excitations, affirming that Nd-Ni hybridization decreases with decreasing dimensionality (*12, 40, 51*).

Together, these observations reinforce the notion, predicted by DFT (Fig. 4B, (*36, 40*)), that hole doping away from a 3$d^9$ configuration, whether through chemical doping or dimensional tuning, enhances the cuprate-like electronic structure in square-planar nickelates (*52*). In this context, it is interesting that over-doped $n = 3$ and 32.5% Sr-doped NdNiO$_2$ do not superconduct. An open challenge then is to identify the features associated with the destruction of superconductivity in the nickelates.

**Magnetic excitations:**

One common candidate for superconducting pairing which arises and disappears with doping is the magnetic interaction within a material. In cuprates, superconductivity emerges upon doping a long-range ordered antiferromagnetic (AFM) phase. In contrast, no long-range AFM order has been observed in underdoped nickelates, though spin fluctuations and signatures of a spin glass phase have been reported in the under- and optimally-doped infinite-layer nickelates (*22, 53*). One question is what happens to these spin fluctuations as the superconducting phase is destroyed with doping. Through structural tuning of $n$ in Nd$_{n+1}$Ni$_n$O$_{2n+2}$ compounds, we can explore the highly over-doped side of the phase diagram beyond what has been achieved with chemical substitution.

Our principal finding is the observation through momentum-resolved RIXS (Fig. 5A, B) of damped magnetic excitations at ~80 meV in superconducting $n = 5$ ($d^{8.8}$) (Fig. 5C-E), which persists past superconductivity well into over-doped, metallic $n = 3$ ($d^{8.67}$) (*36*). These inelastic features are identified as having a magnetic origin through the incident energy and polarization dependence of the RIXS spectra (Figs. S27, S28, (*36*)). Furthermore, the broadness and amplitude of the feature is inconsistent with a phononic origin (Fig. S27, (*36*)). The magnetic features in both $n = 3$ and $n = 5$ compounds can be well-modelled by a single magnetic branch (Fig. 5C) and lack a strong dispersion in momentum along the ($h$, 0) direction (Fig. 5D). Despite the different functional doping, dimensional confinement, and electronic properties of the two compounds, we do not observe notable differences in their excitations other than a slight increase of the damping near the zone boundary in the $n = 3$ compound, consistent with its more metallic nature. These observations are in contrast to chemically-doped infinite-layer nickelates, which have more dispersive features with a dampening and mode energy that change more substantially with doping (*53*). Given the similarity of the spectra between the non-superconducting $n = 3$ and the superconducting $n = 5$ compounds then, the relationship between these magnetic modes and superconductivity is unclear.

One possibility is that the structural form of the $Nd_{n+1}Ni_nO_{2n+2}$ nickelates directly impacts these relatively unchanging spin excitation features. The presence of the fluorite layers, which cause a local lattice expansion of the outermost Ni-O planes, identified by our lattice constant analysis with STEM (Fig. 1B) may reflect changes in the magnetic unit cell (Fig. S24, Table S2) relative to $NdNiO_2$ and give rise to multiple magnon modes $n$ within a single compound, similar to what was observed in bulk $La_4Ni_3O_8$ and $Pr_4Ni_3O_8$ (*54*). These magnon modes are likely all strongly damped and further blurred by the spectrometer resolution, and thus cannot be individually resolved. Consequently, we observe an average of all the $2n$ magnetic branches, with different weights throughout the Brillouin zone (*54*). Hence, we emphasize that ~80 meV represents a lower bound for the magnetic energy scales in this family of materials (*36*). This is comparable to the minimum observed for chemically-doped infinite-layer nickelates and is of similar magnitude to the energy scales observed in many cuprates (*55–57*).

We reiterate that in either scenario, magnetic excitations persist well into the overdoped, non-superconducting side of the nickelate phase diagram. While this persistence is similar to cuprates (*56*), unlike in cuprates, the destruction of superconductivity here is not necessarily associated with a weakening of the magnetic interaction *a priori*: the $n = 3$ compound appears more like superconducting cuprates in oxygen hybridization, quasi-two-dimensionality, and reduced role of the rare-earth band (Fig. 4), though is not superconducting. Future studies attempting to identify the role of magnetism or high hole-doping in nickelates, or to relate $T_c$ to the strength of magnetic interactions, could leverage a combination of chemical doping in multi-layer square-planar nickelates of varying $n$ throughout the superconducting region.

**Discussion and outlook:**

Combined, our results have allowed us to construct a distinct phase diagram for multi-layer square-planar nickelates $Nd_{n+1}Ni_nO_{2n+2}$ (Fig. 2A). The superconducting regime heavily overlaps in nominal electron count with chemically-doped infinite-layer nickelates and hole-doped cuprates, indicating that square-planar nickelate superconductivity occurs over an apparently universal doping range near $3d^9$.

The electronic structure trends are consistent with hole-doping in infinite-layer nickelates and bring lower $n$ layered nickelates closer to the cuprates. As $n$ is decreased, rare-earth $5d$ hybridization decreases and oxygen $2p$ hybridization increases, reaffirming the emergence of cuprate-like properties in lower-dimensional square-planar structures. Furthermore, the unique role of neodymium $4f$ moments in infinite-layer nickelates is reprised in multi-layer $Nd_{n+1}Ni_nO_{2n+2}$ nickelates. The magnetic character of the $4f$ moments alters the superconducting anisotropy contrary to what is expected in a confined thin film geometry. This effect increases with increasing hole doping whether through chemical doping or structural tuning by decreasing $n$, indicating it evolves universally in square-planar nickelates.

We can also note distinctions between the multi-layer square-planar nickelates and their infinite-layer counterparts. Firstly, the effects of neodymium $4f$ moments surprisingly completely wash out the expected enhancement of superconducting anisotropy as $n$ is decreased. This points to a complex interplay between the role of rare-earth magnetism and dimensionality in the superconducting phase of these multi-layer systems, which may be tunable with future realization of superconductivity in $R_{n+1}Ni_nO_{2n+2}$ compounds of different rare-earth cations.

Furthermore, structural tuning allows us to access high doping ranges and hence track the evolution of spin fluctuations deep into the over-doped regime of multi-layer $Nd_{n+1}Ni_nO_{2n+2}$ nickelates. Magnetic excitations persist from superconducting $n = 5$ to highly over-doped, metallic $n = 3$ compounds with minimal modification. This resembles the persistence of magnetic excitations in cuprates through the over-doped regime (56), though the relationship between these magnetic modes and superconductivity is presently unclear.

Another key distinction between the multi-layer nickelates and their infinite-layer counterparts is the structural inequivalence between the individual $n$ $NdNiO_2$ layers as observed by STEM (Fig. 1B, (36)). This may lead to a layer-dependent modulation of other properties, including for example an interpretation of the spin fluctuation dispersion in RIXS as emerging from multiple magnon branches (54, 58). Additionally, in multi-layer cuprates, experimental mapping of the band structure suggests that a variation of properties across multiple layers can explain trends in $T_c$ (59, 60). While this is beyond the scope of our current study, our observed lattice modulation provides evidence toward layer-dependent properties, which may be identified in the future as advancements in STEM-EELS move toward atomic-level resolution of subtle spectroscopic features (61). This also underscores the role of the spacer layer: the steric effect is subtle in the case of $(NdO_2)^-$ fluorite layers but we posit that such spacer layers can be more broadly functionalized. This could involve chemical doping or intercalation to tune structural distortions, which in turn may modify inter-layer couplings, magnetic interactions, chemical pressure, anion coordination, and electronic correlations (51).

Most remarkably, the $d$-filling-dependent superconducting regime of the multi-layer nickelates overlaps with that of the infinite-layer compounds, which apparently unifies the square-planar nickelates across different structural realizations. This is particularly notable considering the structural distinctions, the subtle modifications to magnetic excitations, and even the empirical uncertainties in oxygen stoichiometry (17, 62, 63). Hence, the multi-layer square-planar nickelates give us a new, structurally distinct platform to help distill the ingredients of nickelate superconductivity and examine fundamental theories including for example the role of Kondo coupling (46). Experimentally, we can look toward further exploits of atomic-level design to target new ambient pressure superconducting nickelates. This may include synergizing chemical substitution with structural tuning to enhance $T_c$ or determine its precise functional dependence on hole-doping, and, within the shared $3d^9$ doping range, could even extend beyond the square-planar motif.

**Acknowledgments**: We thank Steve Novakov and John T. Heron for electronic measurement assistance; Donald A. Walko for assistance with surface X-ray diffraction experiments at the Advanced Photon Source; and Gaël Grissonnanche and Brad J. Ramshaw for helpful discussions.

**Funding:** This project was primarily supported by the US Department of Energy, Office of Basic Energy Sciences, Division of Materials Sciences and Engineering, under Award No. DE-SC0021925. Materials growth and electron microscopy were supported in part by the Platform the Accelerated Realization, Analysis, and Discovery of Interface Materials (PARADIM) under NSF Cooperative Agreement DMR-2039380. Electron microscopy was primarily performed at the Cornell Center for Materials Research shared facilities, which are supported by the NSF MRSEC program DMR-1719875. All nanofabrication work was performed at Harvard University's Center for Nanoscale Systems, a member of the National Nanotechnology Coordinated Infrastructure Network supported by NSF award no. ECCS-2025158. The RIXS measurements and their interpretation by S.F.R.T., V.B., M.P.M.D., and M.M. was supported by the U.S. Department of Energy (DOE), Division of Materials Science, under Contract DE-SC0012704. RIXS was performed in part at Diamond Light Source on beamline I21-RIXS under Proposal MM27484. This research used beamline 2-ID of NSLS-II, a US DOE Office of Science User Facility operated for the DOE Office of Science by Brookhaven National Laboratory under contract no. DE-SC0012704. Synchrotron surface X-ray diffraction measurements were performed under beam time award Rapid Access GUP-1013484 at sector 7-ID-C at the Advanced Photon Source, a U.S. DOE Office of Science user facility operated for the DOE Office of Science by Argonne National Laboratory under Contract No. DE-AC02-06CH11357.

G.A.P., D.F.S., and A.Y.J. acknowledge support from the NSF Graduate Research Fellowship Grant DGE-1745303. G.A.P. and A.Y.J. were also supported by the Paul & Daisy Soros Fellowship for New Americans, and A.Y.J. by the Ford Foundation. A.B.T. was supported by DMR-2323970. L. B., B.H.G., and L.F.K. acknowledge support from PARADIM, NSF DMR-2039380. B.H.G. was also supported by the Schmidt Science Fellows in partnership with the Rhodes Trust. J.A.M. acknowledges support from the Packard Foundation, Sloan Foundation, and the Gordon and Betty Moore Foundation's EPiQS Initiative, grant GBMF6760.

**Author contributions:** G.A.P, A.S.B., and J.A.M. conceived of and designed the experiment. G.A.P. and D.F.S. synthesized the samples using oxide MBE with contributions from A.Y.J., Q.S., A.B.T., H.P., C.M.B, and J.A.M. G.A.P. and D.F.S. performed reductions and diffraction. G.A.P. performed transport experiments with contributions from D.F.S. L.B. and B.H.G. performed and analyzed STEM and EELS measurements with support from L.F.K. and D.A.M. S.F.R.T, G.A.P, and D.R.B conducted RIXS experiments and analysis with support from A.N., S.A., K.-J.Z., J.P., V.B., M.P.M.D., and M.M. D.F.S. and H.Z. performed synchrotron X-ray diffraction. H.L. and A.S.B. performed density functional theory calculations. G.A.P. wrote the manuscript with input from all authors. J.A.M. supervised the project.

**Competing interests:** Authors declare that they have no competing interests.

**Data and materials availability:** All data in the main text and the supplementary materials will be available.

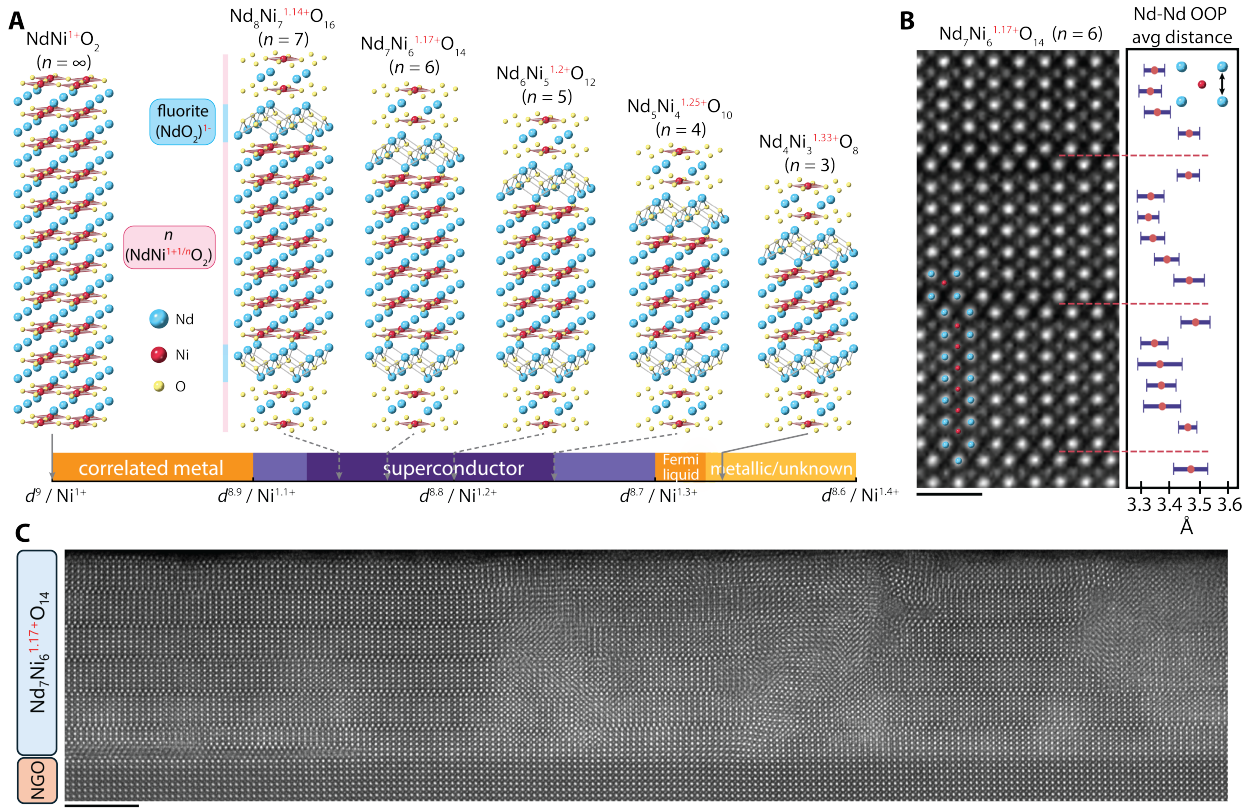

**Fig. 1. Structure of the $Nd_{n+1}Ni_nO_{2n+2}$ compounds.** (**A**) Crystal structure schematic of the $Nd_{n+1}Ni_nO_{2n+2}$ compounds. $n$ repeats of $NdNiO_2$ are sandwiched between charged $(NdO_2)^-$ fluorite layers. The fluorite layers soak up negative charge, nominally redistributing charge across the $NdNiO_2$ layers to $1/n$ holes per nickel. The compounds are mapped onto the doping-dependent phase diagram of the chemically-doped $NdNiO_2$ nickelates. Dark purple range represents the superconducting region taken from initial reports (*26*, *27*); light purple represents the expanded superconducting region from recent improvements (*16*). Dashed lines stand for compounds that cannot be made through bulk synthesis methods. Teal, neodymium; red, nickel; yellow, oxygen. (**B**) HAADF-STEM image and Nd-Nd out-of-plane (OOP) lattice constant analysis of an $n = 6$ block. Dashed red lines indicate $(NdO_2)^-$ fluorite layers. The OOP Nd-Nd distance is larger close to the fluorite layers. Scale bar, 1 nm. (**C**) Large field-of-view STEM image of an $n = 6$ thin film on an $NdGaO_3$ substrate showing largely coherent microstructure. Scale bar, 5 nm.

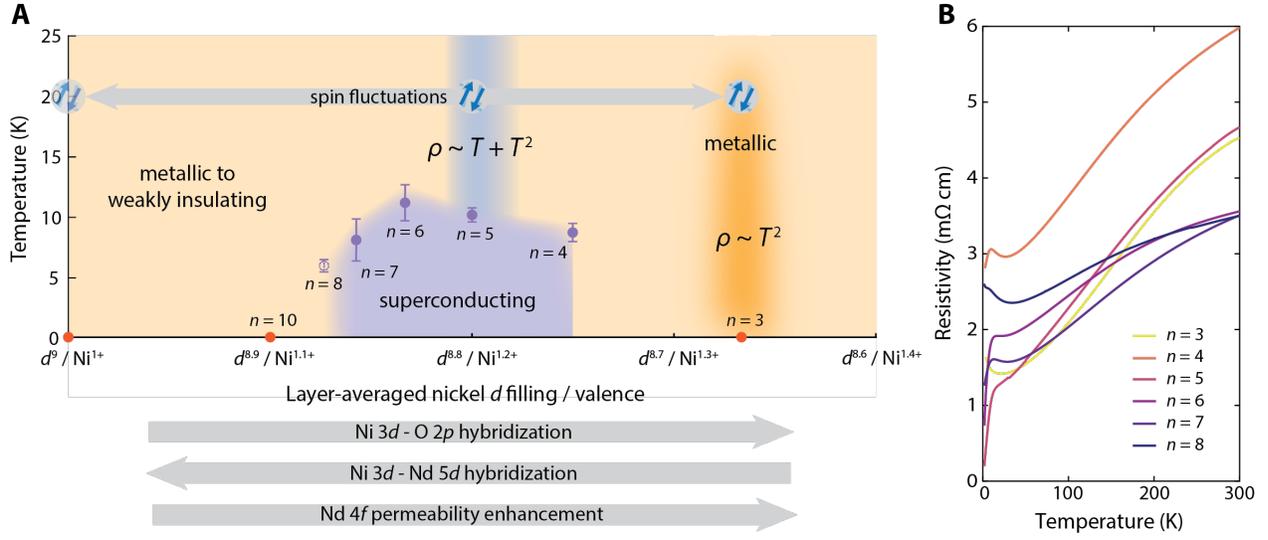

**Fig. 2. Correlated phase diagram of multi-layer $Nd_{n+1}Ni_nO_{2n+2}$ nickelates.** (**A**) Summarized phase diagram of the multi-layer square-planar nickelates $Nd_{n+1}Ni_nO_{2n+2}$ as a function of nominal layer-averaged nickel $d$-electron count. Superconducting phase is in purple; $T_{c,\,onset}$ is determined using the criteria described in Ref. (*36*). Open circle for the $n = 8$ compound represents superconducting correlations without a clear superconducting downturn. Blue arrow symbols represent where spin fluctuations were identified using RIXS; the $d^9$ ($NdNiO_2$) point is taken from Ref. (*53*). Gray arrows (bottom) indicate the evolution of other properties discussed in this manuscript across the phase diagram. $T^2$ and $T$-linear labels are adapted from Ref. (*64*). (**B**) Resistivity of the $Nd_{n+1}Ni_nO_{2n+2}$ compounds for $n = 3 – 8$.

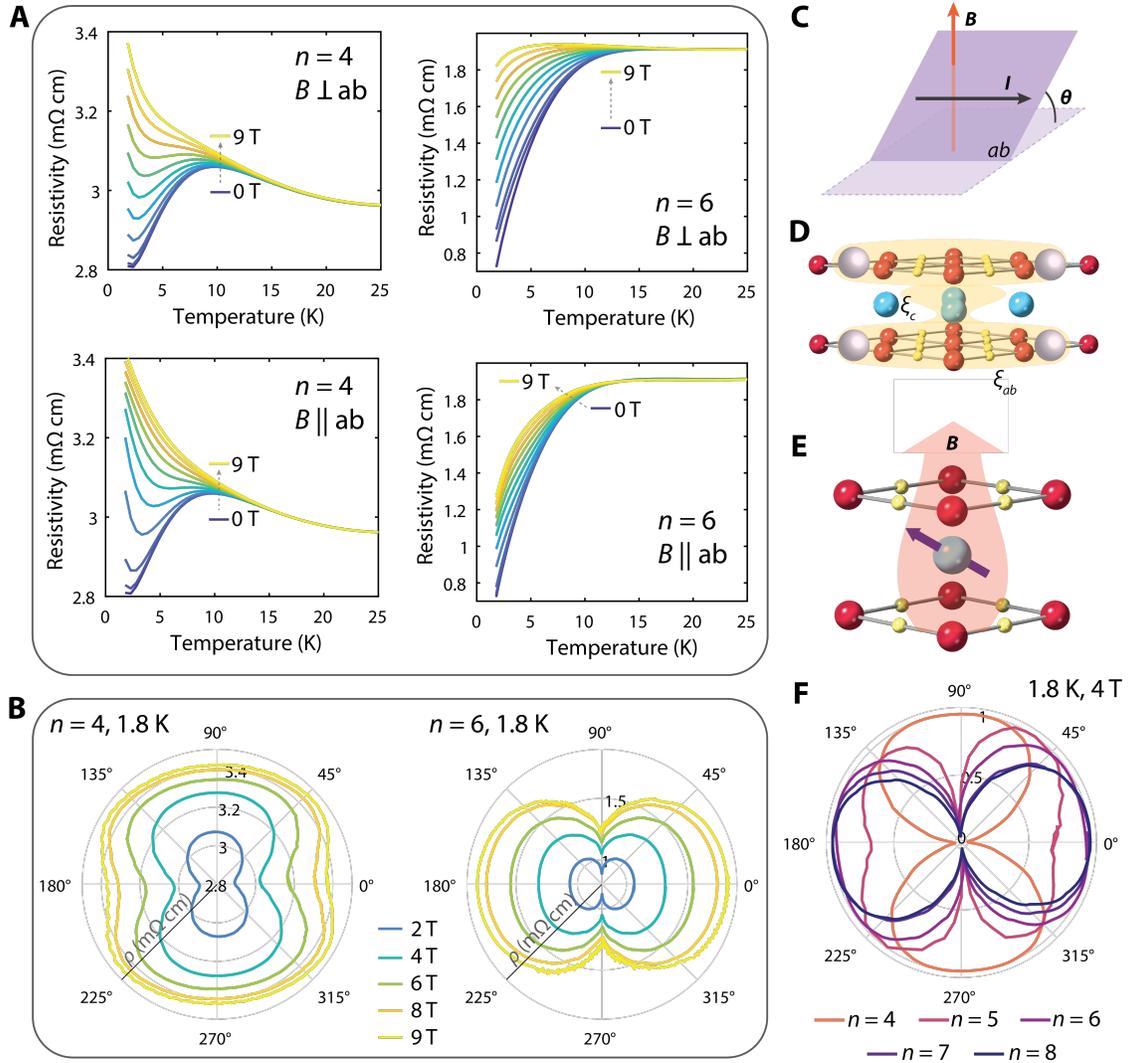

**Fig. 3. Electronic transport characterization of the superconducting phases in $Nd_{n+1}Ni_nO_{2n+2}$.** **(A)** Superconducting transitions of the $n = 4$ and $n = 6$ compounds with applied magnetic fields in-plane or out-of-plane with respect to the sample surface. The $n = 4$ compound (left) appears to behave isotropically in a magnetic field. The $n = 6$ compound (right) shows a clear superconducting anisotropy. **(B)** Angle dependence of the magnetoresistance of the $n = 4$ and $n = 6$ compounds at 1.8 K. **(C)** Schematic of the measurement geometry. **(D)** A "typical" quasi-2D system would show strong anisotropic orbital depairing from different coherence lengths. Gray spheres represent electrons in Cooper pairs. Not to scale. **(E)** A system with 4$f$ moment enhanced magnetic permeability. An external field applied out-of-plane would be internally enhanced (red) along moment directions. **(F)** Angle-dependent magnetoresistance of the $n = 4 – 8$ compounds at 1.8 K and 4 T. Data have been min-max normalized to show all samples on the same scale.

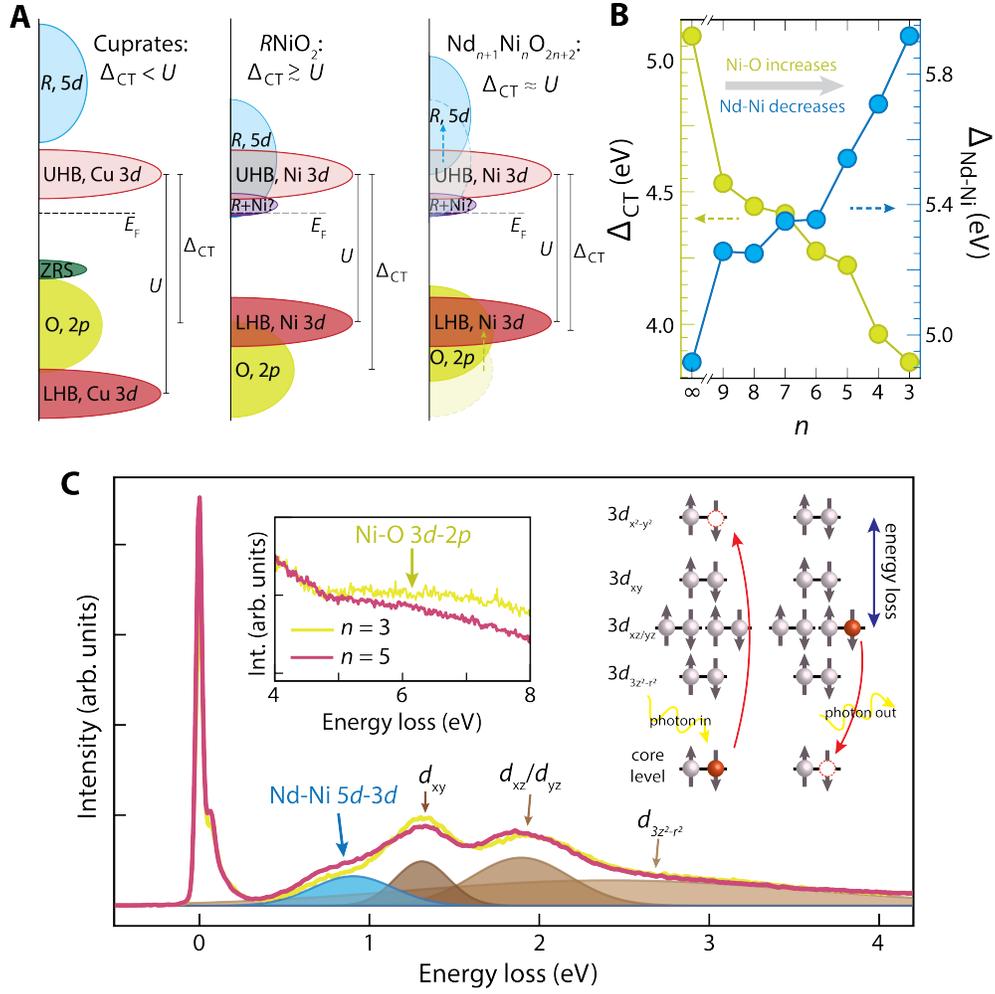

**Fig 4. Electronic structure characterization of the Nd$_{n+1}$Ni$_n$O$_{2n+2}$.** **(A)** Cartoon electronic structure diagrams of the cuprates (left), $R$NiO$_2$ nickelates (center), and Nd$_{n+1}$Ni$_n$O$_{2n+2}$ (right). The dashed lines representing changes in the Nd$_{n+1}$Ni$_n$O$_{2n+2}$ diagram are relative to NdNiO$_2$ as $n$ decreases. $U$, Mott-Hubbard on-site repulsion; $\Delta_{CT}$, charge-transfer gap; LHB, lower Hubbard band; UHB, upper Hubbard band; ZRS, Zhang-Rice singlet. **(B)** DFT predictions, derived from centroids of the atom-resolved density of states (*36*), of the charge-transfer gap and the neodymium-nickel hybridization gap. Increasing gap size indicates decreasing hybridization strength. **(C)** Sample RIXS spectrum of the $n = 3, 5$ compounds at $q_\parallel = (-0.45, 0)$. Orbital excitations are represented by the colored peaks. Left inset: high energy loss feature corresponding to Ni-O hybridization. Right inset: schematic of the RIXS process that generates $dd$-excitations in a square-planar crystal field.

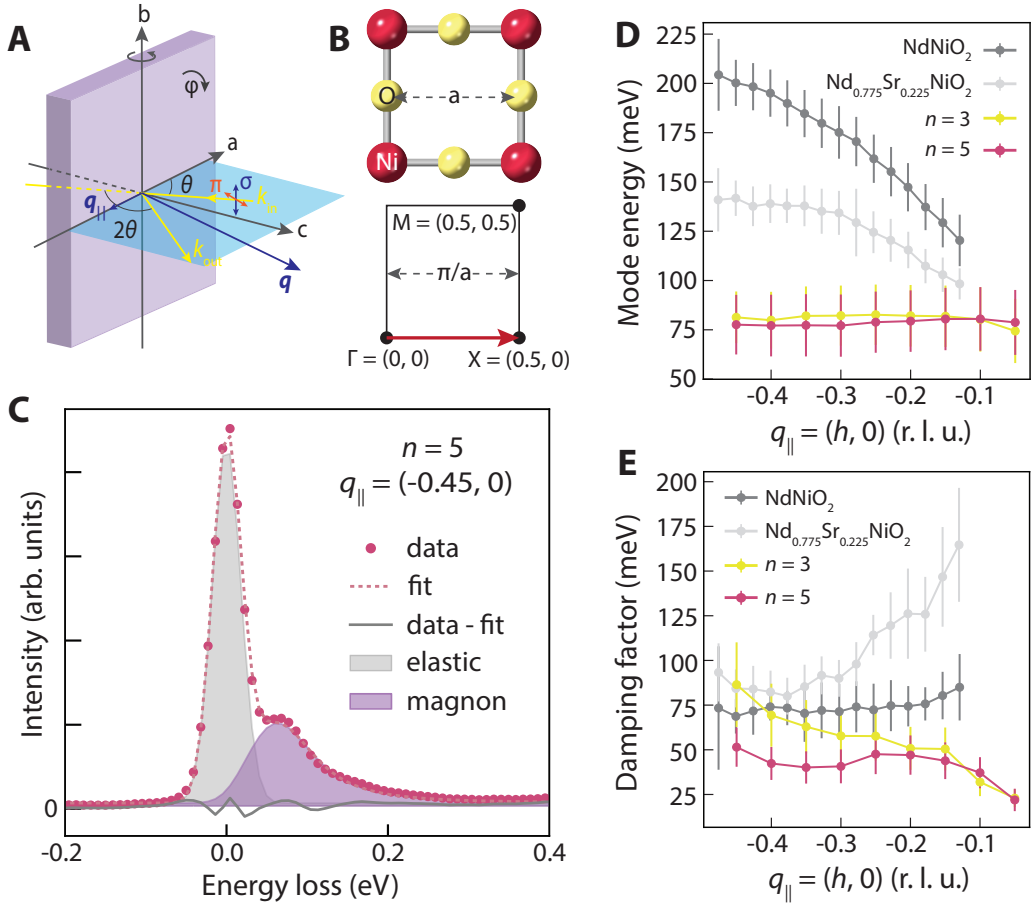

**Fig 5. Magnetic excitations in Nd$_{n+1}$Ni$_n$O$_{2n+2}$ using RIXS.** **(A)** Schematic of the RIXS scattering geometry used to access trajectories in momentum space. Purple represents the sample, blue the scattering plane. **(B)** Real- and momentum-space cartoons of the direction of the momentum scan direction in (D) and (E). **(C)** Example low-energy loss spectrum highlighting the magnon-derived contribution. Energy **(D)** and damping **(E)** terms of the magnetic excitation along the $q_\parallel = (h, 0)$ direction of the $n = 3, 5$ nickelates, compared to infinite-layer nickelates. Infinite-layer nickelate data reprinted from Ref. (*53*).